\documentclass[]{article}
\usepackage{amsmath}
\usepackage{graphicx}
\usepackage{authblk}

\newcommand{\Selfpower}{375}
\newcommand{\RHpower}{9.89}
\newcommand{\COtpower}{1.16}
\newcommand{\SelfRMS}{161}
\newcommand{\RMSRH}{52.4}
\newcommand{\RMSCOt}{5.4}

\newcommand{\annularDNCoeff}{1.3 \times 10^{-19} }

\newcommand{\CITa}{1}
\newcommand{\UFa}{2}
\newcommand{\LHOa}{3}
\newcommand{\LLOa}{4}
\newcommand{\ADLa}{5}

\title{Overview of Advanced LIGO Adaptive Optics}

\author[\CITa]{Aidan F. Brooks}
\author[\CITa]{Benjamin Abbott}
\author[\UFa]{Muzammil A. Arain}
\author[\UFa]{Giacomo Ciani}
\author[\CITa]{Ayodele Cole}
\author[\LHOa]{Greg Grabeel}
\author[\CITa]{Eric Gustafson}
\author[\LLOa]{Chris Guido}
\author[\LLOa]{Matthew Heintze}
\author[\CITa]{Alastair Heptonstall}
\author[\CITa]{Mindy Jacobson} 
\author[\ADLa]{Won Kim}
\author[\ADLa]{Eleanor King}
\author[\CITa]{Alexander Lynch}
\author[\CITa]{Stephen O'Connor}
\author[\ADLa]{David Ottaway}
\author[\CITa]{Ken Mailand}
\author[\UFa]{Guido Mueller}
\author[\ADLa]{Jesper Munch}
\author[\CITa]{Virginio Sannibale}
\author[\CITa]{Zhenhua Shao}
\author[\CITa]{Michael Smith}
\author[\ADLa]{Peter Veitch}
\author[\LHOa]{Thomas Vo}
\author[\LHOa]{Cheryl Vorvick}
\author[\CITa]{Phil Willems} 

\affil[\CITa]{LIGO Laboratory, California Institute of Technology \\ 1200 East California Boulevard, Pasadena, CA 91125, USA}
\affil[\UFa]{University of Florida, Gainesville \\ Florida 32611, USA}
\affil[\LHOa]{LIGO Hanford Observatory, Richland\\ Washington 99352, USA}
\affil[\LLOa]{LIGO Livingston Observatory, Livingston\\ Louisiana 70754, USA}
\affil[\ADLa]{University of Adelaide, Adelaide \\ South Australia 5005, Australia}

\affil[*]{Corresponding author: brooks\_a@ligo.caltech.edu}

\begin{document}

\maketitle

\begin{abstract}
This is an overview of the adaptive optics used in Advanced LIGO (aLIGO), known as the thermal compensation system (TCS). The thermal compensation system was designed to minimize thermally-induced spatial distortions in the interferometer optical modes and to provide some correction for static curvature errors in the core optics of aLIGO. The TCS is comprised of ring heater actuators, spatially tunable CO$_{2}$ laser projectors and Hartmann wavefront sensors. The system meets the requirements of correcting for nominal distortion in Advanced LIGO to a maximum residual error of 5.4nm, weighted across the laser beam, for up to 125W of laser input power into the interferometer.
\end{abstract}

%\setboolean{displaycopyright}{true}

%\begin{document}

%\thispagestyle{fancy}

%\ifthenelse{\boolean{shortarticle}}{\ifthenelse{\boolean{singlecolumn}}{\abscontentformatted}{\abscontent}}{}

%=====================================================================================================
%\input introduction_aLIGO_TCSOSA.tex

% aLIGO TCS introductory section

\section{Introduction: High power advanced gravitational wave detectors}

\subsection{Overview of GW detectors and the search for GW}

The search for gravitational waves has a long history. Recent efforts have focussed on interferometric gravitational wave detectors, such as LIGO, VIRGO, GEO600 \cite{aLIGO_instrument_paper, VIRGOref, GEO_ref_Dooley}. Second generation gravitational wave detectors are now coming online. Indeed, Advanced LIGO (aLIGO) \cite{aLIGO_instrument_paper} recently made the first two direct detections of gravitational waves, GW150914 \cite{GW150914} and GW151226 \cite{GW151226}.

\begin{figure}[htbp]
\centering
\includegraphics[width=10cm]{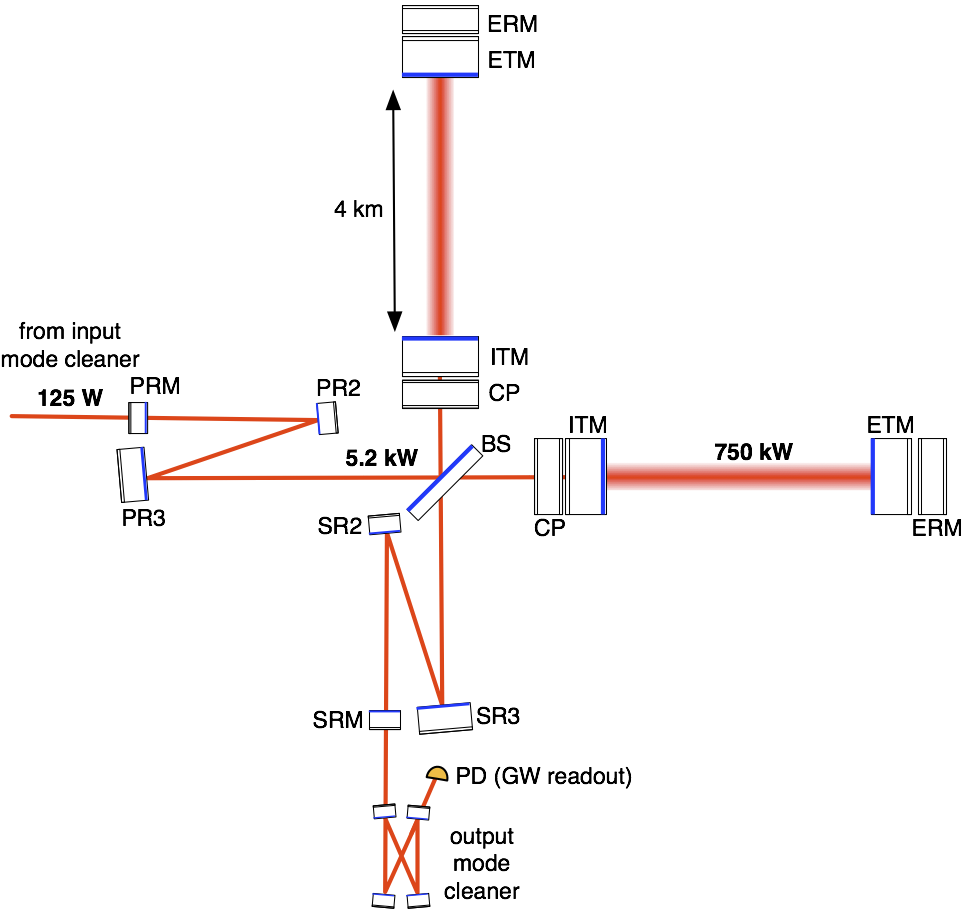}
\caption{aLIGO layout. The pre-stabilized laser beam is injected into an input mode cleaner (not shown). The output of this is injected into the dual-recycled Fabry-Perot Michelson interferometer though mirror PRM. The Fabry-Perot arms are formed between the high-reflectivity (HR) surfaces of the input test mass (ITM) and end test mass (ETM) of each arm where up to 750kW of circulating power is stored. Any differential change in the arm lengths will result in a change in the output signal from the interferometer through the output mirror SRM where it will be spatially filtered by the output mode cleaner and detected on the GW readout photodiode (PD). See reference \cite{aLIGO_instrument_paper}, from which this figure is adapted, for a full description of the layout.} \label{fig:aLIGO}
\end{figure}

Each of the two detectors, Hanford and Livingston, that comprise aLIGO is a dual-recycled Fabry-Perot  Michelson interferometer (IFO), as illustrated in Figure \ref{fig:aLIGO}. Power and signal recycling cavities are installed on the input and output sides of the beamsplitter, respectively. A 1064nm laser beam is resonant in the Fabry-Perot (FP) arms and in the recycling cavities. The strain from a passing gravitational wave incident on the IFO will stretch one arm and shrink the other. This will cause a differential phase modulation on the resonant laser field in the arms resulting in an intensity modulation on the laser field exiting the beam-splitter and signal recycling cavity which is then measured on a photodetector. 

The second generation detectors resemble the Initial \cite{Abbott:2007kv} and Enhanced \cite{Aasi:2014mqd} detectors with a number of key changes to improve the sensitivity of the device ten-fold. These changes are: (a) a signal extraction cavity (colloquially referred to as a signal recycling cavity), (b) improved seismic isolation and suspension of the optics, (c) larger test masses and (d) higher laser power, see \cite{aLIGO_instrument_paper,Fritschel01}. In addition to the main carrier laser field, there are RF modulation sideband fields that are used for length and alignment control and diagnostics. These are resonant only in the recycling cavities.

\begin{figure}[htbp]
\includegraphics[width=10cm]{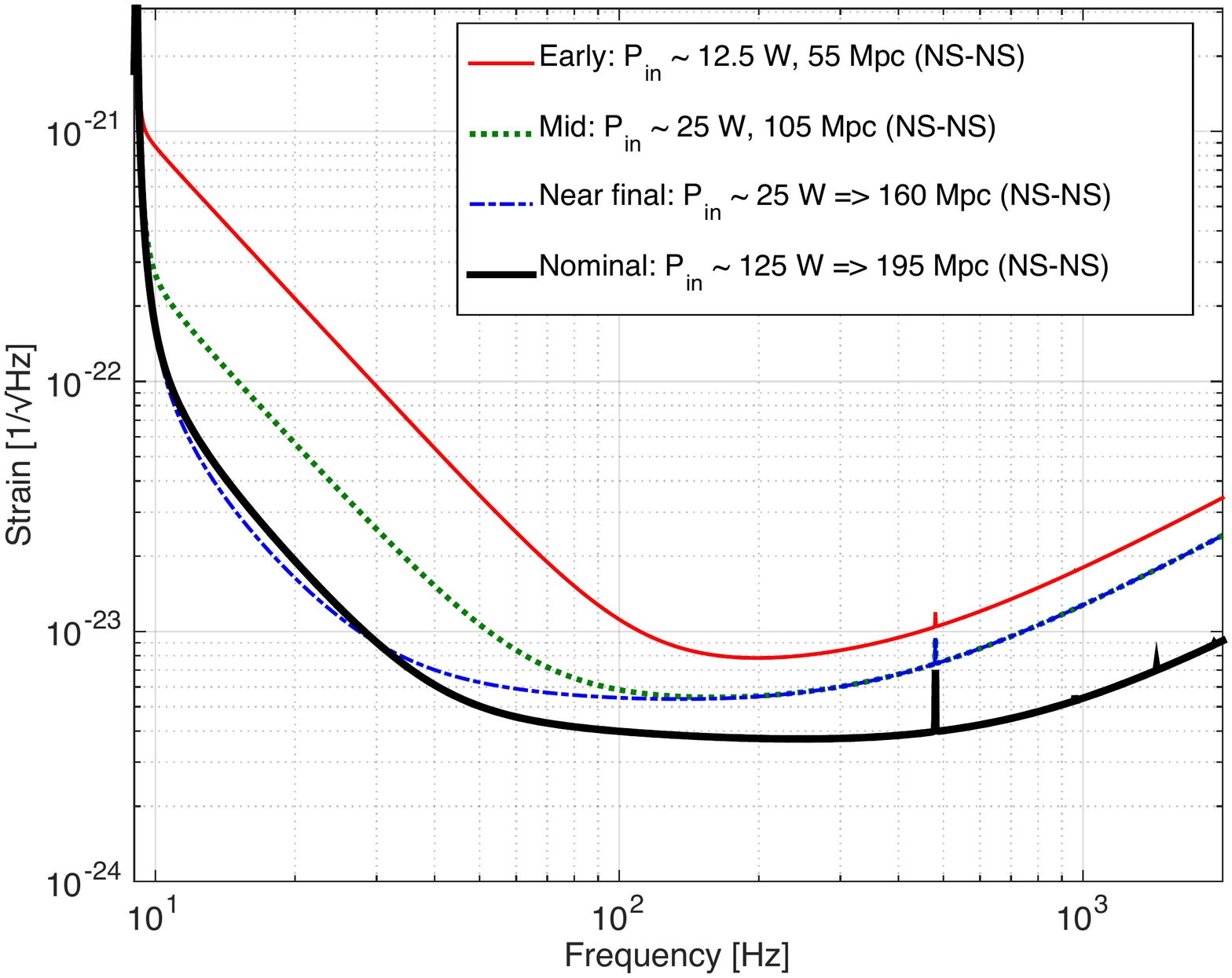}
\caption{Strain sensitivity curves for the different stages of aLIGO. The proposed early configuration (red) is 12.5W input power all the way to 125W input power (to the PRM) for the final aLIGO configuration (black): see reference  \cite{aLIGO_sens}, from which this figure is adapted, for full details.} \label{fig:aLIGO_sens}
\end{figure}

Increasing the stored power in an interferometer decreases the quantum shot noise at high frequencies in the readout, proportional to the inverse of the square root of the stored power.  In aLIGO, we plan to increase the stored power in the interferometer in several stages, as illustrated in Figure \ref{fig:aLIGO_sens}, reproduced from \cite{aLIGO_sens}, from an initial 12.5W input in 2015 to a maximum power of approximately 125W input into the interferometer,  circa 2017. The final configuration will result in approximately 750kW of stored power in the FP arms of the interferometer, yielding an optimum strain sensitivity of  around $4 \times 10^{-23}/ \sqrt{\mathrm{Hz}}$ at 100Hz.
 
Higher stored power in the interferometer results in several effects that make control of the interferometer more difficult, such as radiation-pressure-induced angular instabilities \cite{SidlesSigg}, parametric instabilities \cite{aLIGO_PI2015} and, the focus of this article, thermo-optical distortion  from absorption of optical power \cite{Lawrence_PhD03, Lawrence02}.

%----------------------------------------------------------------

\subsection{Adverse effects of thermo-optical distortion}

\begin{figure}[htbp]
\includegraphics[width=10cm]{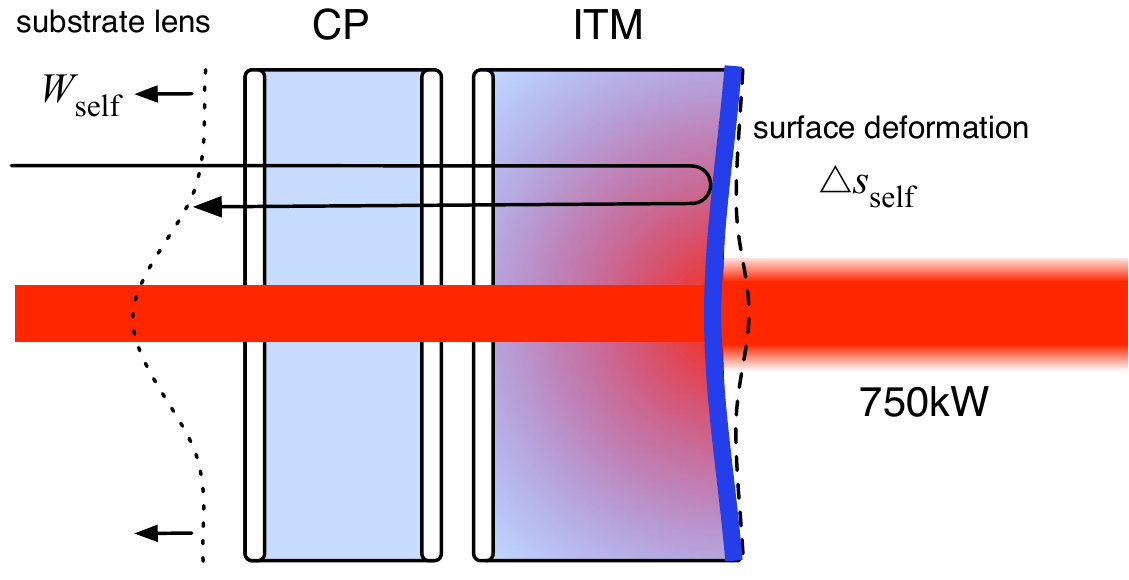}
\caption{An illustration of the thermo-refractive substrate lens, $W_{\mathrm{self}}$, and the thermo-elastic surface deformation, $\Delta s_{\mathrm{self}}$, from self heating.} \label{fig:lensing_overview}
\end{figure}

In aLIGO there will be up to 750kW incident on the surfaces (coatings) of the test masses (ITMs and ETMs in Figure \ref{fig:aLIGO}). Nominal coating absorption is 0.5ppm. As a result, there will be up to \Selfpower mW of laser power with a Gaussian spatial distribution absorbed in the coatings of the test masses. Absorption within the coatings of these optics results in a radial temperature gradient within the optics \cite{Hello90, Winkler91} and, subsequently, thermo-refractive substrate lenses in the recycling cavities  (due to the dependence of refractive index on temperature) and thermo-elastic surface deformation in the Fabry-Perot arms. This is illustrated in Figure \ref{fig:lensing_overview}.

\begin{figure}[htbp]
\includegraphics[width=10cm]{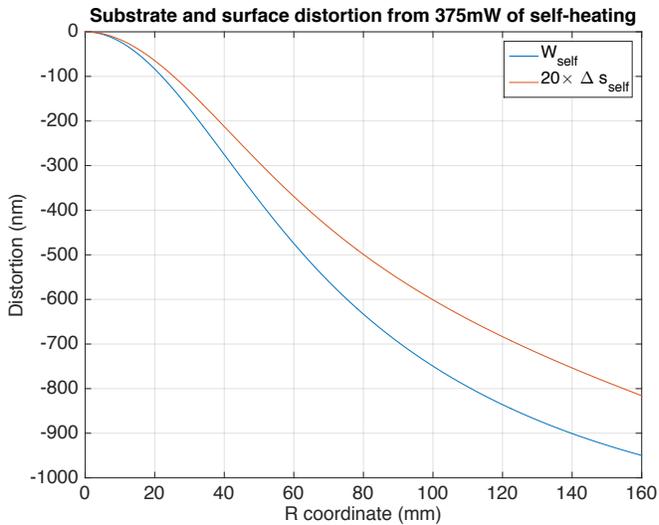}
\caption{Radial distribution of thermo-refractive wavefront distortion, $W_{\mathrm{self}}$, from \Selfpower mW of self-heating. } \label{fig:self_OPD}
\end{figure}

Both phenomena add wavefront distortions to the incident carrier and sideband fields. To distinguish between substrate and surface lenses in this text, we refer to the thermo-refractive lens as wavefront distortion, $W$,  and the surface deformation as a change in the {\it{sagitta}} of the surface, $\Delta s$ - both typically measured in nanometers. The nominal, round-trip, thermo-refractive wavefront distortion from \Selfpower mW of self heating, $W_{\mathrm{self}}$, is shown in Figure \ref{fig:self_OPD} (blue curve). Also shown is the sagitta of the HR surface of the optic, $\Delta s_{\mathrm{self}}$ (scaled by $20 \times$ for clarity).

%From the perspective of optical analysis, we find it convenient to decompose the wavefront distortion into a sum of Zernikes:

%\begin{equation}
%\Delta W = \sum_{i,j} Z_{i,j}
%\end{equation}

The effects of the wavefront distortion on interferometer performance  were  described at length by Ryan Lawrence for Initial LIGO, see \cite{Lawrence_PhD03} for full details. The essential effects are similar for aLIGO and are summarized below. Wavefront distortion causes:

\begin{enumerate}
\item a reduction in the GW signal amplification through mode-mismatch to the Fabry-Perot arms. %Section 2.1 of Ryan Lawrence's thesis.
\item direct (optical) increases in noise at the readout photodiode due to:
\begin{enumerate}
\item a reduction in the power at the beam splitter due to reduced mode-matching into the power-recycling cavity. 
\item `junk light' being reflected toward the anti-symmetric port: differential-mode wavefront distortion in the arms of the interferometer. This `junk light' thus adds shot noise and potentially other technical noise.
\end{enumerate}
\item indirect increases in technical noise from length and alignment control systems due to a reduction in loop gains of those control systems. 
\item a reduction in the stability and robustness of the interferometer due to large wavefront distortions causing severe gain reduction in other control loops.
\item increased coupling of intensity and frequency noise to the DC readout scheme
\end{enumerate}

%TCS shall maintain the extraction efficiency of the gravitational wave sidebands through the signal recycling cavity to the dark port to at least 95% of its nominal value

To minimize these effects, the aLIGO Systems Design specified that the prescribed loss in the extraction efficiency of the gravitational wave sidebands through the signal recycling cavity (SRC) to the dark port shall not exceed 5\%  \cite{Fritschel01, Smith_Willems00}.  Due to the resonant nature of the optical cavities in aLIGO, the maximum allowed 5\% loss of the GW sidebands from the SRC corresponds to approximately 0.1\% loss on a single round-trip through the cavity. 

We represent the carrier field as a sum of transverse electromagnetic-modes (TEM$_{\mathrm{mn}}$), $  \left| U_{mn} \right> $, where the GW sidebands are encoded only in the  $\left| U_{00} \right>$ mode \footnote{For simplicity, these calculations are represented in the Laguerre-Gauss basis, but they can be easily generalized to a Hermite-Gauss basis}:

\begin{equation}
\left| U_{00} \right> =\sqrt{\frac{2}{\pi \, w^2}} \exp\left( - \frac{r^2}{w^2}\right),
\end{equation}

\noindent where $w$ is the Gaussian beam size. Laser power is scattered between modes upon interaction with wavefront distortion. The overlap between two TEM00 modes in the presence of a distortion, $\Delta W$, is:

%\begin{equation}
\begin{multline}
\left< U_{00}  \left| \, \exp\left( i \, k \, \Delta W \right)   \right|  U_{00} \right> = \\ \frac{4}{w^2} \int_{r=0}^{\infty} \exp\left( i \, k \, \Delta W \right) \, r\, \exp\left( -2 \frac{ r^2}{w^2}\right) \mathrm{d}r, 
\end{multline}

\noindent where $k$ is the wavenumber.  Expressed mathematically, the fractional power scattered from the TEM00 mode on one round-trip, $\epsilon_{00}$, is:

\begin{multline}
\epsilon_{00} = 1 - \left< U_{00} \left| \, \exp\left( i \, k \, \Delta W \right) \,  \right| U_{00} \right> \times \\ \left< U_{00} \left| \, \exp\left( - i \, k \, \Delta W \right) \,  \right| U_{00} \right>
\label{eqn:TCS_scatter_M}
\end{multline}

\noindent This round-trip scattering must be less than 0.1\%. A convenient and versatile description is the RMS wavefront distortion,  $\left< \Delta W \right>$, weighted by the intensity distribution of the TEM00 mode:

\begin{equation}
\left< \Delta W \right>^2 = \left< U_{00} \left|  \left( \Delta W - \overline{\Delta W}\right)^2 \,  \right| U_{00} \right>, 
\label{eqn:TCS_scatter_M1}
\end{equation}

\noindent where $ \overline{\Delta W} $ = $\left< U_{00} \left| \,\, \Delta W \,  \right| U_{00} \right>$, which simply appears as a uniform change in phase across the TEM00 mode and is automatically corrected by the length control systems in aLIGO. A purely quadratic wavefront distortion  of approximately 5.4nm RMS will scatter of 0.1\%. This RMS error also serves as a reasonable limit for non-quadratic wavefront distortions that can be formed by heating processes. For reference, $\left<W_{\mathrm{self}}\right>$ for \Selfpower mW absorbed is approximately \SelfRMS nm RMS. 

In aLIGO,  the only place where significant surface deformation and wavefront distortion are created from the self-absorption of interferometer power is on the HR surfaces and within the substrates of the test masses, respectively. In the latter case, only the substrates of the ITMs are within the resonant recycling cavities and present a problem for IFO control. The ETM substrates are outside the IFO and do not cause substantial problems.  An adaptive optical system is required to compensate for the wavefront distortion thermal effects. In aLIGO, this is the Thermal Compensation System (TCS).
%=====================================================================================================

%=====================================================================================================
% TCS components section
%\input aLIGO_TCS_componentsOSA.tex
% aLIGO TCS Components section of paper

\section{Thermal Compensation System (TCS) goals}

The TCS is comprised of a system of actuators and sensors to allow us to compensate for thermal effects in the test masses to reduce the overall distortion on transmission and reflection to less than 5.4nm. The overarching purpose of this whole system is to maintain the interferometer in a given optical configuration such that it may continue to operate reliably. To achieve this the system must:

\begin{enumerate}
\item Measure, with dedicated sensors, the magnitude and spatial distribution of wavefront distortion introduced by self-heating and by any compensation (see Hartmann wavefront sensor in section \ref{sec:HWS}).
\item Compensate for the self-heating induced sagitta change, $\Delta s_{\mathrm{self}}$, by actuating on the radius of curvature (ROC) of the test masses (see Ring Heater in section \ref{sec:RH}).
\item Compensate for the wavefront distortion in the substrate of the ITMs, $W_{\mathrm{self}}$, using wavefront actuators with spatial variability, reducing it to less than 5.4nm RMS (see CO$_{2}$ laser actuator design in section \ref{sec:CO2} \ref{sec:CO2_layout}).
\item Perform all these tasks without injecting additional noise into the GW detector (see Noise in section \ref{sec:CO2} \ref{sec:CO2_noise}).
\end{enumerate}

\begin{figure}[htbp]
\centering
\includegraphics[width=10cm]{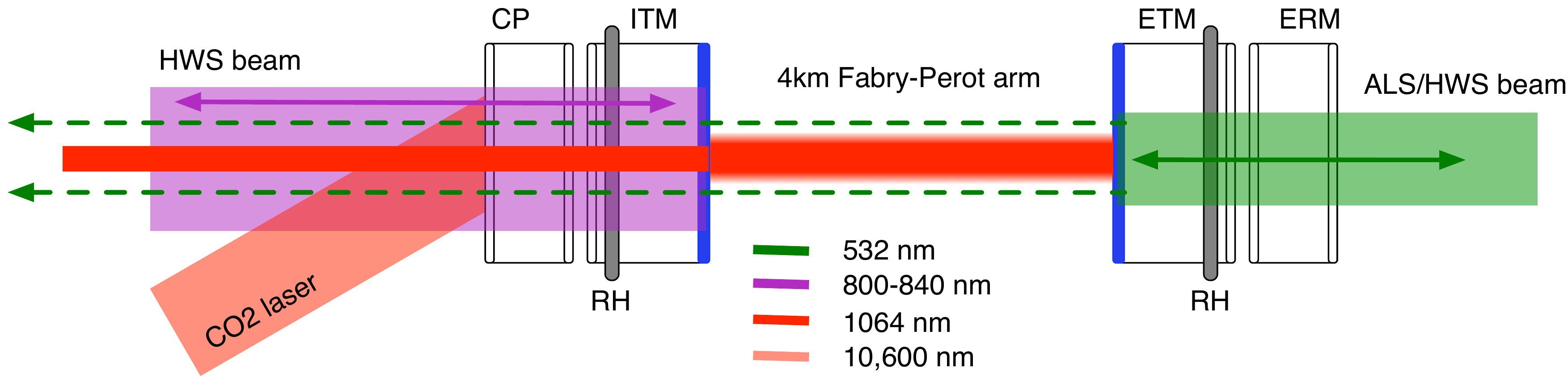}
\caption{aLIGO TCS overview. Absorption of the main interferometer beam [1064nm] (red) in the test masses induces thermal lenses (and subsequent wavefront distortion). Hartmann wavefront sensors (HWS) probe beams, [800-840nm] (purple) and [532nm] (green), measure the thermal lens in the substrates of the ITM+CP and ETM+ERM. Ring heaters (RH), gray, encircle the test masses and radiate  heat onto the barrel to induce thermo-elastic and thermo-refractive distortion. CO$_{2}$ laser beams [10,600nm] (pink) heat the CP to induce a spatially tunable thermal lens. Note that the green laser beam at the ETM is also part of the Arm Length Stabilization (ALS) system \cite{aLIGO_ALS}. A small fraction the green laser leaks through the 4km arm (dashed green) and is used to align the ITM HWS beam to the ITM.} \label{fig:aLIGO_TCS}
\end{figure}

A schematic overview of the TCS sensor beams and actuators used in a single FP arm of aLIGO is shown in Figure \ref{fig:aLIGO_TCS}.  This arrangement is present in both arms. The components are (a) Hartmann wavefront sensors (HWS), using near-IR probe beams [800-840nm] to measure the spatial distribution of the substrate thermal lenses in the ITMs, (b)  Hartmann wavefront sensors (HWS), using visible laser beams [532nm] to measure the spatial distribution of the substrate thermal lenses in the ETMs, (c) ring heater (RH) actuators that heat the barrel of the ITMs and ETMs, altering the surface curvature and substrate lenses of these optics, and (d) CO$_{2}$ laser projectors (10,600 nm) that heat the compensation plate (CP) providing spatially tunable lensing actuation in the recycling cavities. 

The 532nm laser beam used for the ETM HWS is a pick-off from a separate control system in aLIGO, the Arm Length Stabilization (ALS) system \cite{aLIGO_ALS}, and  is aligned to the 4km long Fabry-Perot arm cavity. The leakage field of this beam through the ITMs conveniently identifies the optical axis of the arm and is useful for alignment of the ITM HWS beam.

It is important to note that the Hartmann sensors provide information about the local wavefront distortion in each of the test masses, independent of the resonating 1064nm laser radiation in the interferometer. The baseline design of the TCS does not extract any information about the spatial structure of the interferometer modes that are shaped by local wavefront distortions.

The wavefront distortion induced by self-heating, CO$_{2}$ lasers or RHs may contain many higher spatial aberrations, but it is often convenient to characterize it solely by the {\it{defocus}}, or  quadratic coefficient, $\Delta S$, as this is generally the dominant term in the loss. The wavefront distortion is then approximately:

\begin{equation}
\Delta W \approx \frac{\Delta S}{2} r^2.
\label{eqn:defocus}
\end{equation}

\noindent We rely on the Hartmann sensor to measure the overall defocus induced by the self-heating, the RH and the CO$_{2}$ laser projector. 

%\noindent For self-absorption, decomposition of analytic \cite{Hello90} and finite-element modeling of thermal aberrations into optical aberrations indicate that the single-pass substrate defocus is approximately \SelfDefPerWattGW $\mu$D (micro-diopters) per Watt in the ITM and 386$\mu$D per Watt in the ETM. The defocus of the surface deformation is approximately -37$\mu$D per Watt for the ITM and -29$\mu$D per Watt for the ETM. 

%\section{Components of the aLIGO TCS}
%===========================================================================
%\input HWS_section.tex
%=============================================================================
\section{Hartmann wavefront sensor}
\label{sec:HWS}

\subsection{HWS design}

\begin{figure}[htbp]
\centering
\includegraphics[width=10cm]{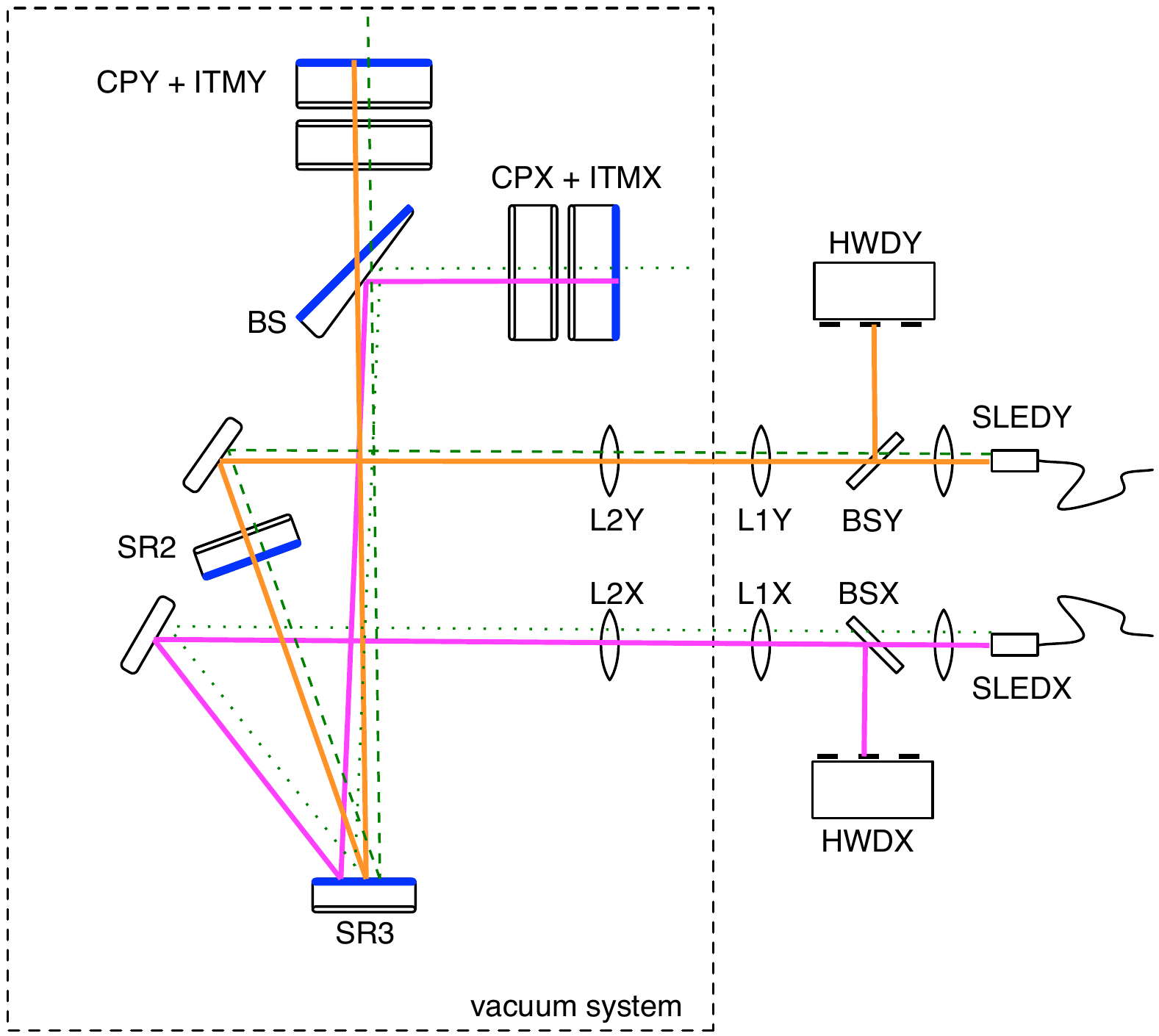}
\caption{The optical layout of the ITMX and ITMY Hartmann wavefront sensors. Probe beams (X-arm, purple, Y-arm, orange) are produced by fiber-coupled super-luminescent diodes (SLEDX, SLEDY). After being collimated, these beams are passed through several imaging optics inside and outside the vacuum system. The beams are then steered into the interferometer core optics (SR2, SR3) and up to the beam splitter (BS - shown with an exaggerated wedge) where the X-arm beam is reflected from the anti-reflecting (at 1064nm) side of the BS and the Y-arm beam is transmitted. Each beam passes through its respective compensation plate and ITM substrate and is retro-reflected from the ITM HR surface. The beams return all the way out of the vacuum system where they are picked off by 50/50 beam splitters (BSX, BSY) and are incident on their respective detectors (HWD). Leakage of the two green ALS lasers beams (shown by dotted and dashed green lines) are used to align the HWS beams to the test masses \cite{HWS_align}.} \label{fig:aLIGO_HWS}
\end{figure}

For brevity, only the design of the ITM HWS is described in detail. The ETM HWS are conceptually very similar, but with a different probe beam wavelength. We refer to the whole Hartmann wavefront sensor system as HWS, while the Hartmann wavefront detector at the core of the system is referred to by the acronym HWD.

The ITM Hartmann wavefront sensors, illustrated in Figure \ref{fig:aLIGO_HWS}, are designed to measure the total substrate thermal lenses seen by the IFO optical modes in the recycling cavities.  The performance requirements on the HWS, driven by the requirement to scatter no more than 0.1\% of the TEM00 mode of the interferometer, are summarized in Table \ref{tab:aLIGO_HWS_req}.

      \begin{table}
      \centering
      \begin{tabular}{ l  c }
       \hline   \bf{Parameter}  &  \bf{Requirement}\\  \hline
       Wavefront sensitivity & 1.35 nm  \cite{Willems06_sensor_req}\\
       Minimum spatial resolution & 1cm $\times $ 1cm  \cite{Willems06_sensor_req}\\
       Spatial extent & $\approx$ 200 mm diameter \cite{TCS_FDD} \\ 
       Measurement frequency & $\ge 200$ mHz  \cite{TCS_FDD} \\
       \hline
      \end{tabular}
      \caption{The HWS requirements for aLIGO.}
       \label{tab:aLIGO_HWS_req}
      \end{table}

%\begin{itemize}
%\item a wavefront sensitivity of at least 1.35nm ,
%\item a spatial resolution of approximately 1cm x 1cm at the surface of the test mass  \cite{Willems06_sensor_req},
%\item probe a region of the test mass that covers at least 99.9\% of power in the interferometer TEM00 beam width \cite{TCS_FDD}, and 
%\item a measurement frequency of at least 200mHz, fast enough to observe the transient thermal lenses in the test mass \cite{TCS_FDD}.
%\end{itemize}

The Hartmann wavefront detector (HWD) is based on a design from the University of Adelaide \cite{Brooks07}. As illustrated in Figure \ref{fig:aLIGO_HWS}, we inject collimated probe beams from fiber-coupled super-luminescent diodes. The X-arm probe is a broad-band 2.5mW, 40nm line-width beam, centered around 800nm (selected to minimize interference effects). The Y-arm probe is also 2.5mW, 18nm line-width beam, centered around 833nm \cite{HWS_SLEDs}. The beams are expanded and collimated by imaging optics (L1X and L1Y) and transmitted through a window into the vacuum system. Inside the vacuum system, they are passed through several more imaging and alignment optics until they are injected into the interferometer. 

The X-arm beam passes through SR3, is reflected off the "anti-reflecting" (AR) surface of the BS (the surface is AR for the interferometer wavelength, not the HWS beam) and finally passed through the substrate of the CP and ITMX. There it is retro-reflected from ITMX HR and is returned out of the vacuum system. 

The Y-arm beam is injected through the rear surface of SR2, sent to SR3, through the BS and is passed through the substrate of the CP and ITMY. There it is reflected off the HR surface of ITMY and returned out of the vacuum system. Alignment of the in-air optics to the in-vacuum optics is a simple matter of matching the in-going HWS probe beams to the axis of the out-going ALS green beams, shown by the overlapping dashed green beams.

Once the beams are returned out of the vacuum system, 50/50 beam splitters are used to extract the return beams and direct them onto the HWDs.

The imaging systems of both the X-arm and Y-arm are designed to image the HR surface of the ITM onto the surface of the detectors with a magnification of 1/17.5 \cite{HWS_ITM_optical}, such that a region of 200mm on the ITM just encompasses the full 12mm diameter of the HWD CCD, provisionally satisfying the requirement on spatial extent. The wavefront is sampled every 430$\mu$m on the HWD, corresponding to sampling every 7.5mm at the ITM, satisfying the requirement on spatial resolution. 

Intensity noise on the HWS probe beam exerts a force, via radiation pressure, on the ITM and, as such, couples to displacement noise in the interferometer. Calculations indicate that this will be at least 400$\times$ lower than the best aLIGO displacement noise floor \cite{HWS_DN}.

There are a few differences between the ETM and ITM HWSs. Rather than a dedicated light source, the ETM HWS uses a pick-off of the green ALS beam used for auxiliary length control of the arm cavity \cite{aLIGO_ALS}. Due to integration with the ALS system, the ETM is imaged onto the ETM-HWD  with a magnification of 1/20 \cite{HWS_ETM_optical}. From an operational perspective, the ETM HWS are used as occasional diagnostic tool, whereas the ITM HWS are  run continuously during operation.

\subsection{HWS operation}

\begin{figure}[htbp]
\centering
\includegraphics[width=10cm]{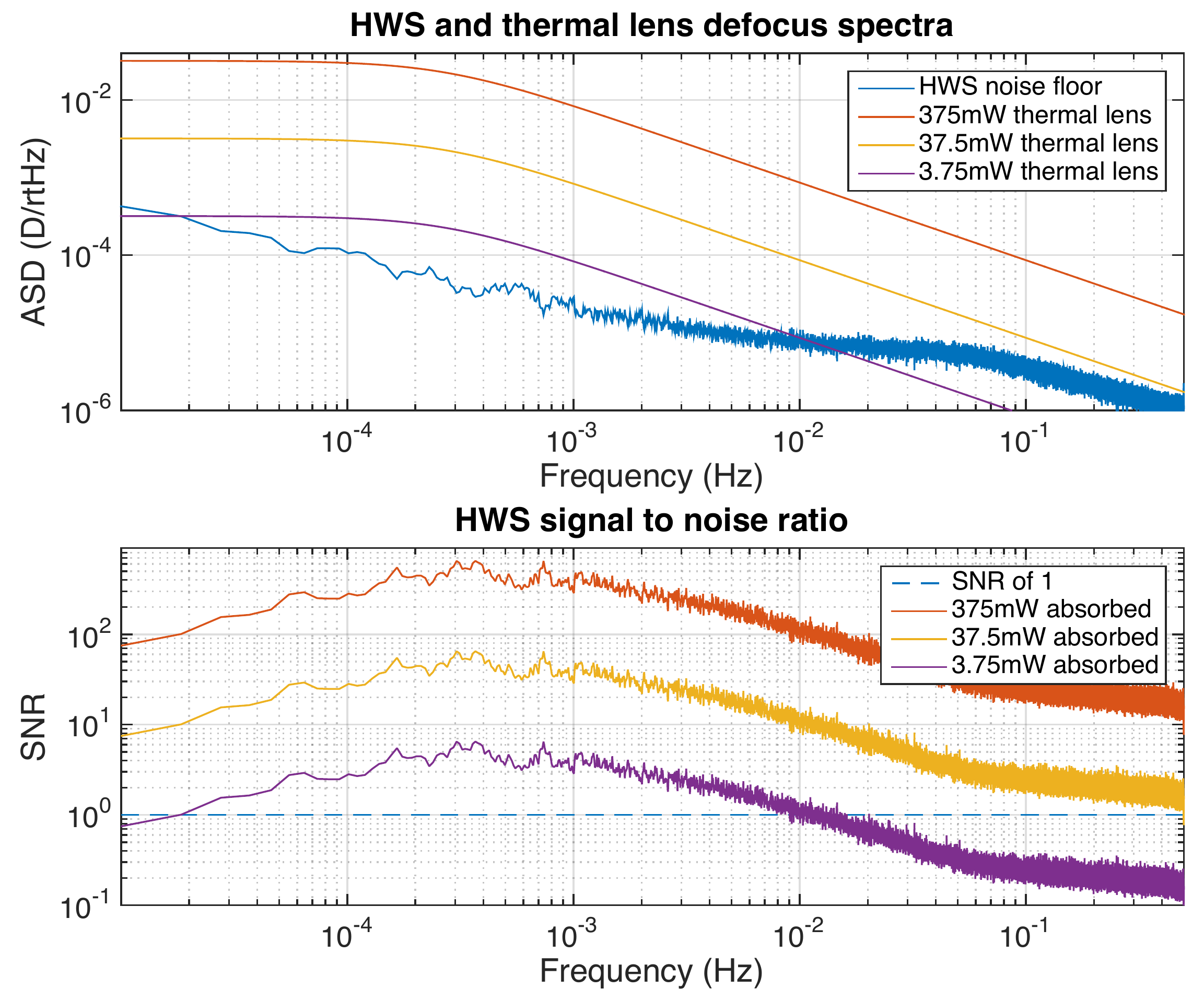}
\caption{Background defocus noise spectra (in diopters) of the aLIGO HWS. The expected thermal lens magnitudes for different frequencies and absorbed powers are also shown.} \label{fig:aLIGO_HWS_meas}
\end{figure}

The HWS measures changes in the gradient, $\nabla$, of the accumulated wavefront distortion, $\Delta W_{\mathrm{HWS}}$ relative to some reference state  \cite{Brooks07}. This measured gradient is decomposed into  the Seidel optical aberrations that represent the wavefront \cite{HWS_SAD}, including the predominant one: defocus, $S$.

Due to shot noise, jitter in the HWS probe beam, air currents on the in-air table, temperature fluctuations and long-term drift in the optics, there is a background error in the measurement of the wavefront distortion, $W_{\epsilon}$, that has some frequency dependence. As a result, the sensitivity of the HWS is best on a time scale of around 3000s. Thus, the most precise measurements are made by resetting the reference wavefront immediately prior to any dedicated measurements. During self-heating, for example, we would reset the reference immediately before the interferometer is locked.  The observed background noise in the defocus measured by the aLIGO HWS is shown in Figure \ref{fig:aLIGO_HWS_meas}. The figure illustrates that the HWS is able to measure the thermal lens with an absorbed power of 3.75mW with an SNR of approximately 5 when measured over 3000s. The required wavefront sensitivity required 1.35nm corresponds to approximately 1.8$\mu$D, or around 3.5mW of absorbed power and the HWS meets this requirement. 

%By combining,  Equation  \ref{eqn:TCS_scatter_M1}  and Equation \ref{eqn:defocus} and the wavefront sensitivity requirement of 1.35nm, we find that the maximum allowed defocus error of the HWS is determined to be approximately 1.9 $\mu$diopters (for a interferometer TEM00 mode,  $\left| U_{00} \right>$, with a beam size of 54mm (???)).

% data is in /Advanced_LIGO/LLO_thermal_lens/HWS_noise

%\input RH_section.tex
%----------------------------------------------------------------
% Ring heater

\section{Ring heater actuator}
\label{sec:RH}

\subsection{RH design}

The primary TCS actuator is the ring heaters (RH):  a glass torus, wrapped in nichrome wire through which current is dissipated to heat the barrel of the optic it encircles. Each test mass has a RH actuating on it. The primary goal of a RH is to compensate for self-heating induced sagitta change, $\Delta s_{\mathrm{self}}$, by actuating on the ROC of its test mass. It can also be used to tune the g-factor and higher order mode spacing of the FP arms to assist in the control of parametric instabilities \cite{aLIGO_PI2015}.

\begin{figure}[htbp]
\centering
\includegraphics[width=10cm]{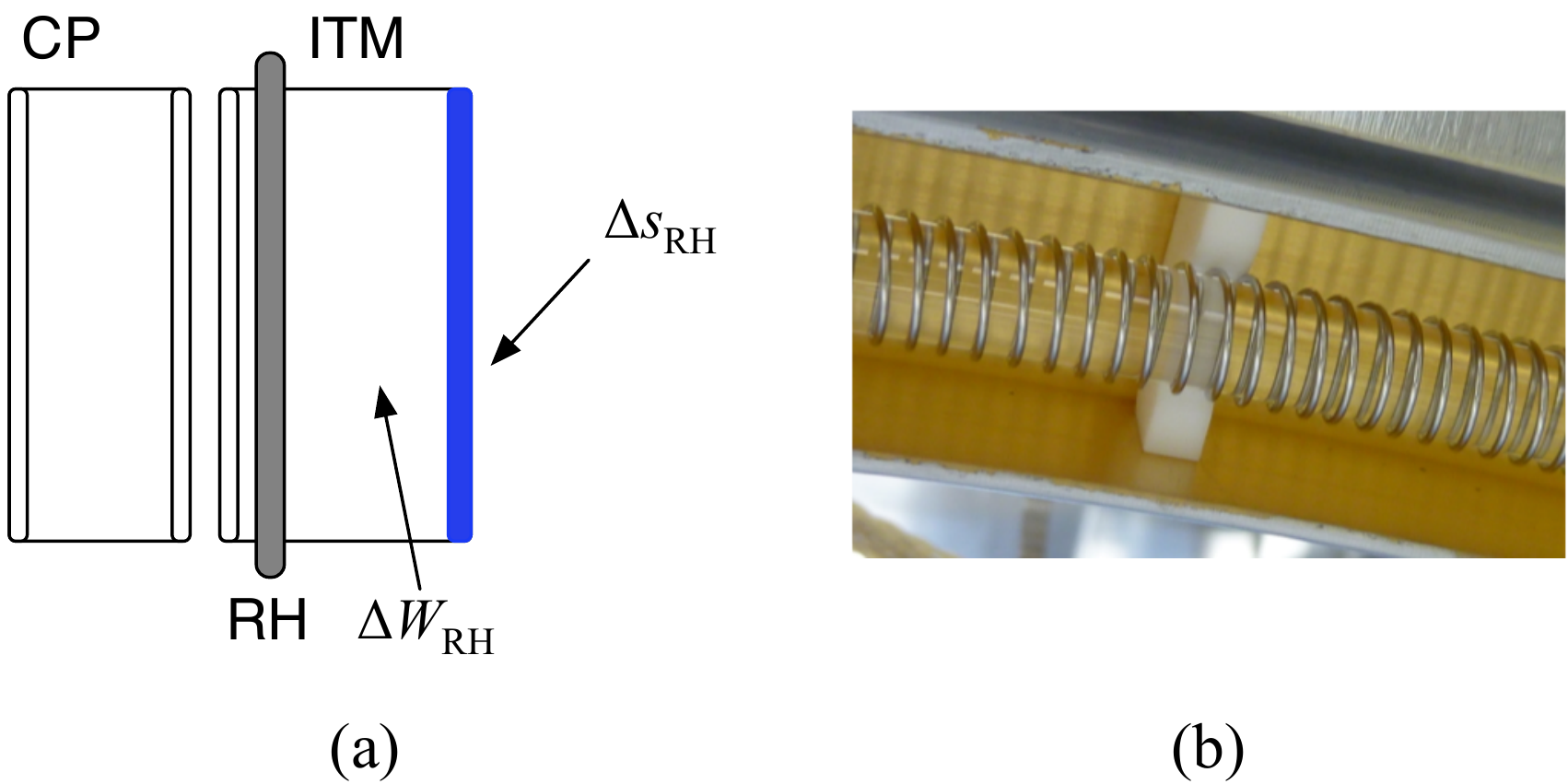}
\caption{(a) the aLIGO RH torus, (b) the nichrome wire encircled RH torus} \label{fig:RH_figure}
\end{figure}

In detail, a RH is a shielded glass torus, 175mm major radius, 3mm minor radius,  that encircles the barrel of each of the test masses, see Figure \ref{fig:RH_figure}(a). The torus is constructed from two semicircular segments that are attached to the cage of the quad suspension system \cite{aLIGO_instrument_paper}. The glass segments are wrapped with nichrome wire, a section of which is shown in Figure \ref{fig:RH_figure}(b). Electrical current is passed through the wire segments, heating them and the glass. The glass then radiates energy onto the test mass barrel.  Each segment is driven by an independent current driver. The shield that surrounds the outer half of the segments is gold coated to maximize heat reflected from it onto the barrel. The segments are held in place in the shield by small ceramic elements (one of which is shown in \ref{fig:RH_figure}(b)).

\begin{figure}[htbp]
\centering
\includegraphics[width=10cm]{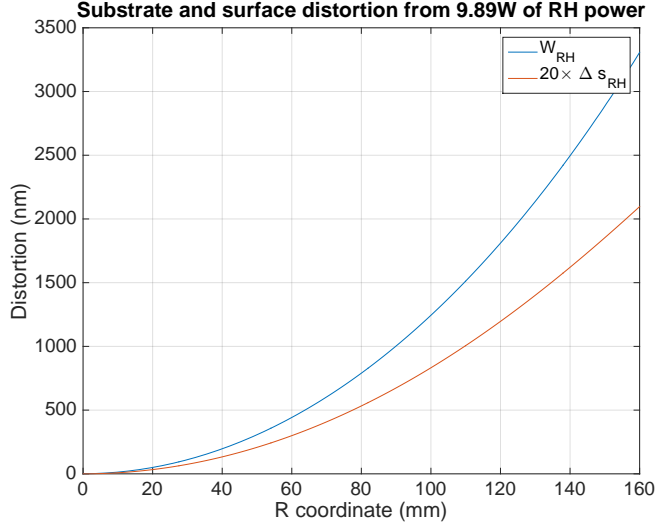}
\caption{The round-trip wavefront distortion from \RHpower W  of RH power, $W_{\mathrm{RH}}$.}
\label{fig:RH_OPD}
\end{figure}

When operating, the RH creates a radial temperature gradient within the test mass. This, in turn, adds positive radial sagitta, $\Delta s_{\mathrm{RH}}$, to the front surface (making it more concave). Additionally, the temperature gradient creates a negative  thermo-refractive lens, $W_{\mathrm{RH}}$, in the substrate. Both $\Delta s_{\mathrm{RH}}$ and $W_{\mathrm{RH}}$ are almost entirely quadratic and are illustrated in Figure \ref{fig:RH_OPD}. 

The ROC of the front surface of the test mass is reduced (making it more curved). An analytic model of the RH thermal lens is available here \cite{RHmodel2016}. The steady-state defocus change, $\Delta S_{\mathrm{HR}}$, is linearly proportional to the change in RH power, $\Delta P$, multiplied by the coefficient, $\Delta S/\Delta P$, approximately $1.0\mu$D per watt. The corresponding change in the ROC of a test mass is given by:

\begin{equation}
\Delta R = -\Delta S_{\mathrm{HR}} \, R^2
\end{equation}

Each current source is capable of producing up to 20W of power per segment, for a total maximum power  of 40W for a RH. For the ITM, with a ROC of 1934m, this corresponds to a maximum $\Delta R$ of approximately 150m. The ETM has a ROC of 2240m and corresponding maximum $\Delta R$ of 200m.

The optimum RH power to correct $\Delta s_{\mathrm{self}}$ from \Selfpower mW of heating  is approximately \RHpower W (a RH thus has plenty of range in power to meet this requirement). This creates $\Delta s_{\mathrm{RH}}$ shown in Figure \ref{fig:RH_OPD}. Additionally, $W_{\mathrm{RH}}$ is produced in the substrate. The RMS wavefront error of $W_{\mathrm{RH}} + W_{\mathrm{self}}$ is \RMSRH nm.

\subsection{Deviations from quadratic distortion}

\begin{figure}[htbp]
\includegraphics[width=10cm]{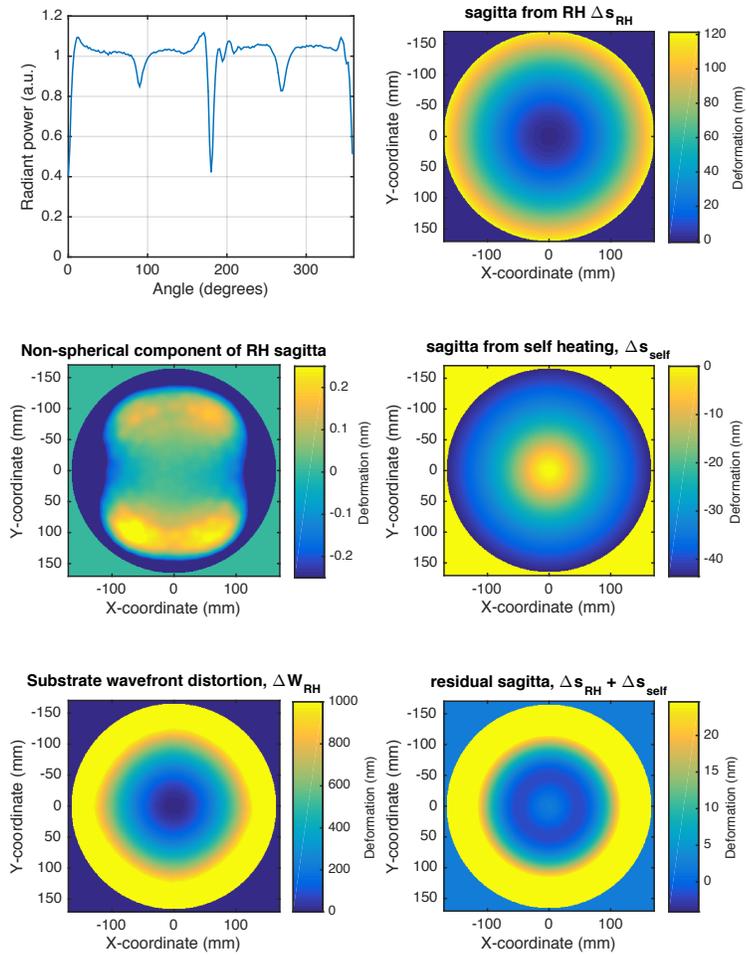}
\caption{Example of the residual thermal lens error (once spherical power has been removed) from the non-idealized RH}
\label{fig:RH_residual}
\end{figure}

Ideally, a RH would provide 100\% axially symmetric heat to the optic. Unfortunately,  engineering design considerations, such as requiring the ability to remove the RH without removing the test mass, prevented this and the RH consists of two semi-circular segements. Each segment is held in place with small ceramic pieces at the ends of that segment. These  provide conductive paths for heat to escape the RH and yield a slight drop in the radiated power from the ends of the segment. This in turn causes a small, roughly astigmatic deviation in the heat distribution around the barrel. We mitigate this effect somewhat by increasing the nichrome wire coil density near the ends of the RH segments, increasing the local temperature \cite{RH_mod_astig}. An example of of the radiated heat from an aLIGO RH is shown in Figure \ref{fig:RH_residual} (top left panel). When this heat distribution is modeled in a finite element model of the test mass, we find a large quadratic surface deformation, Figure \ref{fig:RH_residual} (top right panel), a small amount of astigmatism in the surface deformation, Figure \ref{fig:RH_residual} (center left panel) and a slightly astigmatic substrate thermal lens,  Figure \ref{fig:RH_residual} (bottom left panel). When the deformation due to \Selfpower mW  of self-heating is calculated, Figure \ref{fig:RH_residual} (center right panel), and added to the RH correction, we find the residual deformation shown in Figure \ref{fig:RH_residual} (bottom right panel). This has an RMS value of approximately 1.6nm. In this simulation, the RH has a total output power of 10.24W.

%\begin{figure}
%\centerline{\includegraphics[width=10cm]{RH_radiant_power.pdf}}
%\caption{Example of the radiant power distribution from the RH} \label{fig:RH_radiant_power}
%\end{figure}

\subsection{RH operation}

There are several time constants associated with the operation of the ring heater. When a RH is turned on at time, $t = 0$, it takes approximately 15 minutes for the glass torus of the RH to reach steady-state temperature. The defocus of the surface curvature and substrate thermal lens increase and peaks at a maximum value around $t = 2.7$ hours. Finally, there is an approximately 24 hour time constant for the test mass to reach a steady-state thermal lens of approximately 53\% the size of the peak thermal lens. These time constants make the RH impractical for correcting transient effects such as those associated with a lock-loss of the interferometer. Rather, through an iterative series of adjustments based on measurements from the Hartmann sensor and from other interferometer signals (e.g. contrast defect), optimum RH power settings are chosen for a given IFO operating power and the RHs are fixed at these power levels regardless of the transient behavior of the IFO. 

Voltage noise in the RH will couple to displacement noise in the test mass via electro-static force noise. Provided that the voltage noise is less than approximately 10mV/$\sqrt{\mathrm{Hz}}$ between 10Hz and 100Hz, the RH noise coupling is negligible  \cite{TCS_noise_coupling}. This is very easily achieved with the RH drivers.

%----------------------------
% CO$_{2}$ laser

\section{CO$_{2}$ laser actuator}
\label{sec:CO2}
After the self-heating and ring heater corrections, there will be residual higher-spatial frequency wavefront distortion, $W_{\mathrm{RH}} + W_{\mathrm{self}}$, to correct. Compensating this residual is the job of the second TCS actuator: the CO$_{2}$ laser projector.

\subsection{Layout of the projector}
\label{sec:CO2_layout}

The CO$_{2}$ laser projector applies a spatially tunable heat distribution to the compensation plates. In this way, thermal lenses with virtually any spatial distribution can be created in the recycling cavities.

\begin{figure}[htbp]
\centering
\includegraphics[width=10cm]{{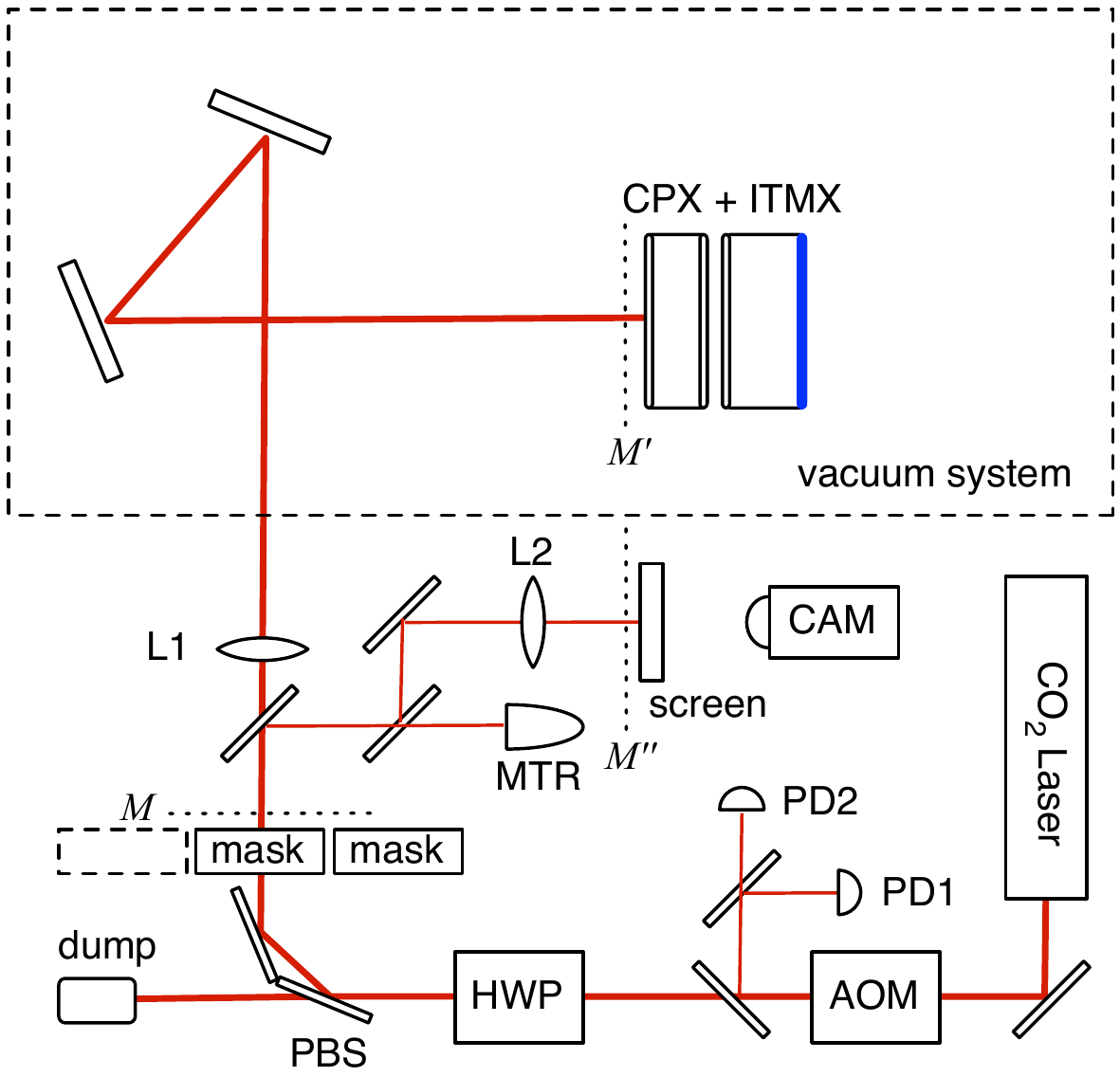}}
\caption{Schematic layout of CO$_{2}$ laser projector} \label{fig:CO2_laser_schematic}
\end{figure}

The CO$_{2}$ laser projector, illustrated in Figure \ref{fig:CO2_laser_schematic}, is the most complex part of the TCS in aLIGO, containing several key components. The first is a temperature and power stabilized CO$_{2}$ laser capable of producing a 50W of TEM00 laser beam with a wavelength of 10.6$\mu$m. The laser beam  passes through an acousto-optic modulator (AOM) that provides high bandwidth, low range, intensity control  of the CO$_{2}$ laser and allows amplitude modulation of the intensity in the audio band. Then the beam passes through a DC power control stage consisting of a half-wave plate (HWP) mounted in a motorized rotation stage and two polarizing beam splitters (PBSs). The beam is magnified by a telescope (not shown) to a Gaussian beam radius of 10mm and it passes through a binary beam shaping mask. After beam shaping, the beam passes through an imaging lens (L1) and into the vacuum system where it is ultimately incident on the BS side of the CP. The imaging is set up such that the CP is a conjugate plane, $M'$ of the mask plane, $M$, with a magnification of $21\times$.

Additionally, there are beam pick-offs for controls and diagnostics.  A pick off after the AOM sends a small amount of light onto two HgCdTe photodiodes for monitoring the intensity noise of the CO$_{2}$ laser in the audio band. When combined with the AOM these sensors allow for audio-band intensity stabilization (approximately 10Hz - 200Hz).  After the beam shaping mask there is a second pick-off, the transmission of which is immediately split into two beams. The first beam is incident on a power meter (MTR) to monitor the power delivered to the test mass. The second beam is incident on a screen that is at a conjugate plane,  $M''$, of the test mass and the mask. This screen is viewed by a far-IR imaging camera (CAM) to monitor the intensity distribution of the laser radiation on the test mass.

Alignment of the CO$_{2}$ laser to the test mass is achieved using a temporarily co-aligned visible laser (for coarse alignment) and measurements of the induced thermal lens with the Hartmann sensor (for fine alignment). 

\subsection{Spatial control of wavefront distortion}

The beam shaping masks  are capable of being exchanged for different designs and heat patterns. 
At any one time, one of two masks can be inserted into the beam by remote controlled flipper mirrors, giving the option for two different heating beam distributions. Further distributions could be applied by manually replacing one of the mask with a new one.
The nominal design includes a central heating mask that provides a negative radial temperature gradient (positive lensing) and an annular heating mask that provides, roughly, negative lensing. Central heating is generally used to compensate for small static lenses  in the substrates of the ITMs induced during manufacturing and to minimize transient thermal effects during down time in the interferometer. Annular heating is used to compensate for the residual distortion $W_{\mathrm{RH}} + W_{\mathrm{self}}$.

The masks are designed by iterating different designs in a  finite-element model of the test mass until the model converges to a design that has the minimal RMS wavefront error, small overlap with the interferometer mode, minimum total required power and is a physically machinable pattern \cite{CO2_mask_construction}.

\begin{figure}[htbp]
\includegraphics[width=10cm]{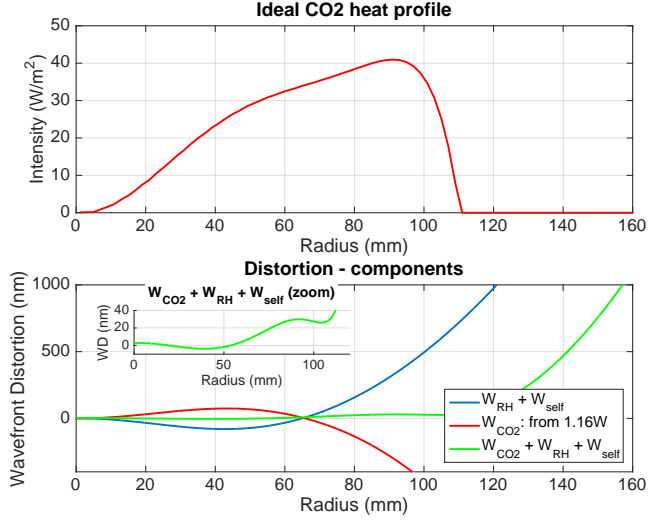}
\caption{(upper) Optimum CO$_{2}$ heating pattern and (lower) the resulting wavefront distortion due to thermo-refractive lensing.} 
\label{fig:CO2_WD}
\end{figure}

For this review of the TCS aLIGO design, it suffices to analyze the masks required to correct for uniform absorption (the same methodology can be applied to non-uniform absorption and custom mask designs). In this case, we must correct for the round-trip wavefront distortion after the RH and self-absorption have been applied, $W_{\mathrm{RH}} + W_{\mathrm{self}}$. At full power, we calculate the correction pattern shown in Figure \ref{fig:CO2_WD} (upper), which contains a total of \COtpower W of power. The resulting wavefront error is shown in Figure \ref{fig:CO2_WD} (lower). The RMS of the residual distortion on transmission through the substrate lens, $\left< W_{\mathrm{CO2}} + W_{\mathrm{RH}} + W_{\mathrm{self}} \right>$ is approximately \RMSCOt nm - just within the requirements.

Corrections for the RH astigmatism and for the fact that the CO$_{2}$ laser has an angle of incidence of approximately 8 degrees on the CP (introducing approximately 2\% ellipticity) can be achieved by non-axially-symmetric perturbations to the nominal mask design.

The time constant of the  thermal lens induced by  the CO$_{2}$ laser varies depending on the spatial distribution. As the CO$_{2}$ laser beam extends to roughly twice the size of the self-heating region, it is of the order of 4$\times$ the time constant of that self-heating thermal lens. This ratio is small enough to make CO$_{2}$ laser heating practical for correcting self-heating problems in real-time.

 \subsection{Noise}
 \label{sec:CO2_noise}
 
\begin{figure}[htbp]
\centering
\includegraphics[width=10cm]{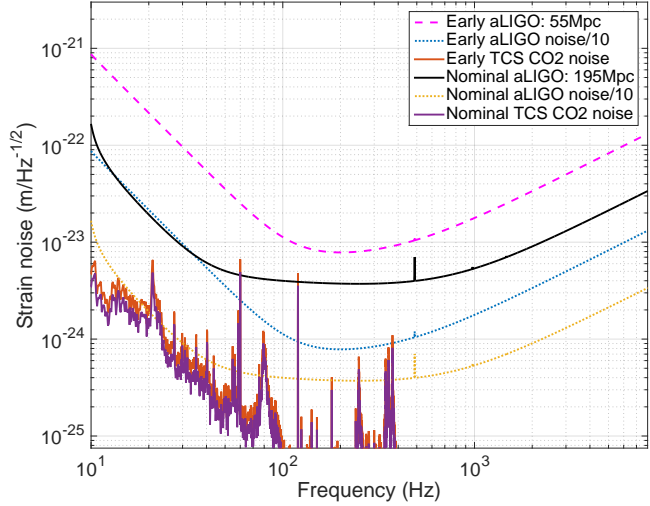}
\caption{Estimated displacement noise from the CO$_{2}$ laser for LIGO during Early and Nominal aLIGO. The dotted blue line is the maximum allowed technical displacement noise for Early aLIGO. The solid red line is the estimated contribution of displacement noise from approximately 750mW of central heating. The dotted yellow line is the maximum allowed technical displacement noise for Nominal aLIGO. The solid purple  line is the estimated contribution from approximately \COtpower W of annular heating.} 
\label{fig:CO2_RIN_DN}
\end{figure}

 Temporal fluctuations in the CO$_{2}$ laser intensity create fluctuations in the optical thickness of the compensation plates, both due to thermo-refractive and thermo-elastic effects which appears as displacement noise in the IFO, potentially becoming a dominant noise source \cite{CO2_noise_coupling}. Indeed, we calculate that the free-running noise of the laser, containing some acoustic peaks around 20Hz, 80Hz and 300-400Hz, could be a dominant noise source if annular heating were used with \COtpower W of heating applied to the test mass, as illustrated in Figure \ref{fig:CO2_RIN_DN}. As such, the intensity can be stabilized by a factor of approximately 10$\times$ to eliminate this effect. However, this is only expected to be necessary in several years time when the interferometer is running at full power.

As described in  \cite{CO2_noise_coupling}, the coupling from laser intensity noise to displacement noise (or strain noise) is a function of the overlap integral of the CO$_{2}$ intensity distribution with the intensity distribution of the interferometer mode - more or less spatial overlap increases or decreases the coupling, respectively. Using the parameters and analysis from \cite{CO2_noise_coupling}, we determine the intensity to strain noise coupling, $\left<h_{CO2}\right>$, for the nominal CO$_{2}$ heat distribution at  \COtpower W shown in Figure \ref{fig:CO2_WD} (upper):

\begin{equation}
\left<h_{CO2}\right> = \annularDNCoeff \left( \frac{100\, \mathrm{Hz}}{f} \right) \mathrm{RIN}(f),
\end{equation}

\noindent where RIN(f) is the relative intensity noise of the CO$_{2}$ laser. Examples of predicted CO$_{2}$ strain noise spectra are shown in Figure \ref{fig:CO2_RIN_DN} for central heating (different coupling coefficient) and the nominal CO$_{2}$ heat distribution shown in Figure \ref{fig:CO2_WD} (upper). These show that, in some instances, a small amount of intensity stabilization may be required.

% get sensitivity curves

%=====================================================================================================

%% TCS measurements
%\input aLIGO_TCS_measurements.tex

% TCS and aLIGO Operation in the future
%\input aLIGO_TCS_future_operation.tex

% TCS and aLIGO operation in the future

\section{Conclusion}

\begin{figure}[htbp]
\centering
\includegraphics[width=10cm]{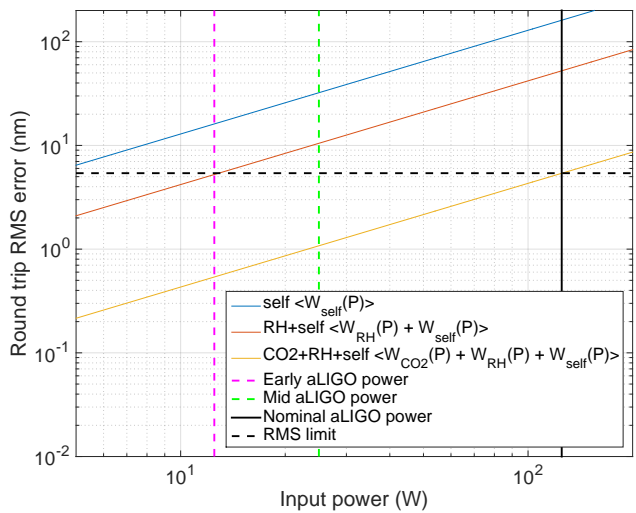}
\caption{RMS error vs aLIGO input power. Also shown are the operating power levels for aLIGO. The maximum allowed RMS is 5.4nm, shown by the black dashed line.} 
\label{fig:TCS_WD_v_P1}
\end{figure}

\begin{figure}[htbp]
\centering
\includegraphics[width=10cm]{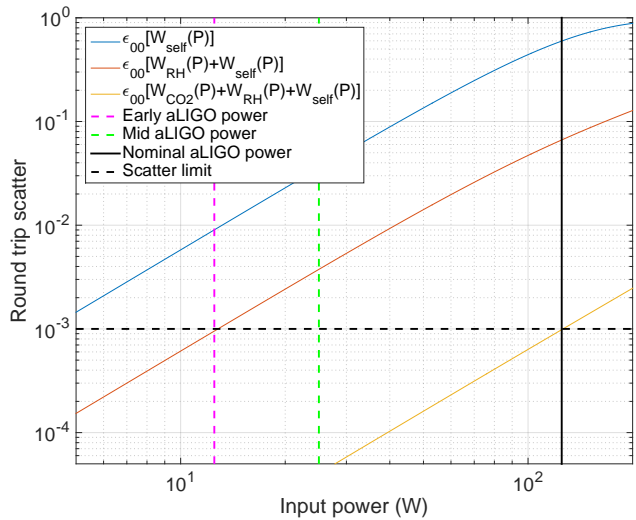}
\caption{Round-trip scatter vs aLIGO input power. Also shown are the operating power levels for aLIGO. The maximum scatter 0.1\%, shown by the black dashed line.} 
\label{fig:TCS_WD_v_P2}
\end{figure}

The operation plans for aLIGO call for several stages of successively increasing power level. The responses of all the TCS actuators scale linearly with interferometer power, as does the wavefront distortion induced by self-heating. As a result, the residual RMS wavefront error increases linearly as a function of power. Figures  \ref{fig:TCS_WD_v_P1} and \ref{fig:TCS_WD_v_P2}  show the predicted residual wavefront error and scatter, respectively, for no-correction ($W_{self}$), only RH correction ($W_{RH} + W_{self}$), and  nominal correction ($W_{CO2} + W_{RH} + W_{self}$) as a function of input laser power. The design nominal correction just meets the scatter limit of 0.1\% for full input power of 125W.

The Thermal Compensation System described in this document allows us to sense and correct for wavefront distortion in aLIGO up to an input laser power of 125W into the IFO. This covers the full proposed operational range of the aLIGO GW detector.

%A summary of the wavefront distortion from the design is presented in Table \ref{tab:aLIGO_TCS_summary}. At the time of writing, several measurements of wavefront error were available and it seemed informative to include the,

     % \begin{table}
     % \centering
     % \begin{tabular}{ l  c c}
      % \hline   \bf{Region}  &  \bf{RMS wavefront error} & \bf{Scatter}\\  \hline \hline
       % \bf{Surface deformation (thermo-elastic)}  & \\\hline
       %Total allowed & 4.564 nm & 0.1\%\\
       %Uncompensated (0.5ppm, 750kW) & 20 nm & 1.5\%\\ 
      % Residual (with 11 W RH) & 1.83 nm & 100 ppm   \\
       %\hline
        %\bf{Substrate lens (thermo-refractive)}  & \\\hline
       %Total allowed & 4.564 nm \\
      % Uncompensated (0.5ppm, 750kW) & 240 nm ??\\ 
      % Residual (with 11 W RH \& 4.5W annular CO2) & 1.83 nm & 100 ppm  \\
       %\hline
      %\end{tabular}
      %\label{tab:aLIGO_TCS_summary}
      %\caption{The wavefront distortion levels in the interferometer}
      %\end{table}

%\section*{Appendix}
%\input aLIGO_TCS_appendix.tex

\section*{Acknowledgements}
The authors gratefully acknowledge the support of the United States
National Science Foundation (NSF) for the construction and operation of the
LIGO Laboratory and Advanced LIGO as well as the Science and Technology Facilities Council (STFC) of the
United Kingdom, the Max-Planck-Society (MPS), and the State of
Niedersachsen/Germany for support of the construction of Advanced LIGO 
and construction and operation of the GEO600 detector. Additional support for Advanced LIGO and this work in particular was provided by the Australian Research Council. LIGO was constructed by the California Institute of Technology and Massachusetts Institute of Technology with funding from the National Science Foundation, and operates under cooperative agreement PHY-0757058. Advanced LIGO was built under award PHY-0823459.

This article has been assigned LIGO document number LIGO-P1600169.

% Full bibliography added automatically for Optics Letters submissions
% Note that this extra page will not count against page length

%Manual citation list
%\begin{thebibliography}{1}
%\bibitem{Zhang:14}
%Y.~Zhang, S.~Qiao, L.~Sun, Q.~W. Shi, W.~Huang, %L.~Li, and Z.~Yang,
 % \enquote{Photoinduced active terahertz metamaterials with nanostructured
  %vanadium dioxide film deposited by sol-gel method,} Opt. Express \textbf{22},
  %11070--11078 (2014).
%\end{thebibliography}


\begin{thebibliography}{10}
\newcommand{\enquote}[1]{``#1''}

\bibitem{aLIGO_instrument_paper}
{The LIGO Scientific Collaboration}, \enquote{Advanced {LIGO},} Class. Quantum
  Grav. \textbf{32}, 074001 (2015).

\bibitem{VIRGOref}
{F. Acernese et al}, \enquote{{Advanced Virgo: a second-generation
  interferometric gravitational wave detector},} {Classical and Quantum
  Gravity} \textbf{32}, 024001 (2015).

\bibitem{GEO_ref_Dooley}
{Kate Dooley, et. al.}, \enquote{{GEO 600 and the GEO-HF upgrade program:
  successes and challenges},} {Classical and Quantum Gravity}  (2015).

\bibitem{GW150914}
{B.P. Abbott, et al.}, \enquote{{Observation of Gravitational Waves from a
  Binary Black Hole Merger},} {Phys. Rev. Lett.} \textbf{116}, 061102 (2016).

\bibitem{GW151226}
{B.P. Abbott, et al.}, \enquote{{GW151226: Observation of Gravitational Waves
  from a 22-Solar-Mass Binary Black Hole Coalescence},} {Phys. Rev. Lett.}
  \textbf{116}, 241103 (2016).

\bibitem{Abbott:2007kv}
B.~P. Abbott \emph{et~al.}, \enquote{{LIGO: The Laser interferometer
  gravitational-wave observatory},} Rept. Prog. Phys. \textbf{72}, 076901
  (2009).

\bibitem{Aasi:2014mqd}
J.~Aasi \emph{et~al.}, \enquote{{Characterization of the LIGO detectors during
  their sixth science run},} Class. Quant. Grav. \textbf{32}, 115012 (2015).

\bibitem{Fritschel01}
P.~Fritschel, \enquote{Advanced {LIGO} systems design,} Tech. Rep.
  T010075-00-D, LIGO Project, https://dcc.ligo.org/LIGO-T010075/public (2001).

\bibitem{aLIGO_sens}
L.~Barsotti and P.~Fritschel, \enquote{{Early aLIGO Configurations: example
  scenarios toward design sensitivity},} Technical Note T1200307-v4, LIGO
  (2012).

\bibitem{SidlesSigg}
J.~A. Sidles and D.~Sigg, \enquote{Optical torques in suspended {Fabry-Perot}
  interferometers,} Phys. Lett. A \textbf{354}, 167--172 (2006).

\bibitem{aLIGO_PI2015}
{M. Evans et. al.}, \enquote{Observation of parametric instability in {Advanced
  LIGO},} Phys. Rev. Lett. \textbf{114}, 161102 (2015).

\bibitem{Lawrence_PhD03}
R.~Lawrence, \enquote{Active wavefront correction in laser interferometric
  gravitational wave detectors,} Ph.D. thesis, Massachusetts Institute of
  Technology (2003).

\bibitem{Lawrence02}
{R. Lawrence, M. Zucker, P. Fritschel, P. Marfuta and D. Shoemaker},
  \enquote{{Adaptive Thermal Compensation of Test Masses in Advanced {LIGO}},}
  Classical Quant. Grav. \textbf{19}, 1803--1812 (2002).

\bibitem{Hello90}
P.~Hello and J.-Y. Vinet, \enquote{Analytical models of thermal aberrations in
  massive mirrors heated by high power laser beams,} J. Phys.-Paris
  \textbf{51}, 1267--1282 (1990).

\bibitem{Winkler91}
W.~Winkler, K.~Danzmann, A.~R{\"u}diger, and R.~Schilling, \enquote{Heating by
  optical absorption and the performance of interferometric gravitational-wave
  detectors,} Phys. Rev. A \textbf{44}, 7022--7036 (1991).

\bibitem{Smith_Willems00}
M.~Smith and P.~Willems, \enquote{Auxiliary optics support system design
  requirements document, vol. 1: Thermal compensation system,} Technical Note
  T000092-02, LIGO (2000). {https://dcc.ligo.org/LIGO-T000092/public}.

\bibitem{aLIGO_ALS}
{A. Staley et. al}, \enquote{Achieving resonance in the advanced ligo
  gravitational-wave interferometer,} Class. Quantum Grav. \textbf{31}, 245010
  (2014).

\bibitem{HWS_align}
{A. Brooks et. al.}, \enquote{{Vertex Hartmann Sensor: Initial and Maintenance
  Alignment Procedures},} Technical Note T1100149-v15, LIGO (2014).

\bibitem{Willems06_sensor_req}
P.~Willems, \enquote{Estimate of {TCS} sensor requirements,} Technical Note
  T060068-v2, LIGO (2006). {https://dcc.ligo.org/LIGO-T060068/public}.

\bibitem{TCS_FDD}
{A. Brooks et al.}, \enquote{{Thermal Compensation System (TCS): Hartmann
  Wavefront Sensor (HWS): Final Design Document},} Technical Note T1100517-v7,
  LIGO (2012).

\bibitem{Brooks07}
A.~F. Brooks, T.-L. Kelly, P.~J. Veitch, and J.~Munch, \enquote{Ultra-sensitive
  wavefront measurement using a {Hartmann} sensor,} Opt. Express \textbf{15},
  10370--10375 (2007).

\bibitem{HWS_SLEDs}
A.~Brooks, \enquote{{Technical Note for the aLIGO TCS Hartmann Sensor Camera
  and Sources},} Technical Note T1000682-v6, LIGO (2011).

\bibitem{HWS_ITM_optical}
{A. Brooks et. al.}, \enquote{{aLIGO Hartmann Sensor Optical Solution
  descriptions (H1, L1) Input Test Masses},} Technical Note T1000179-v17, LIGO
  (2013).

\bibitem{HWS_DN}
A.~Brooks, \enquote{{Hartmann sensor noise to displacement noise coupling in
  aLIGO},} Technical Note T1100518-v2, LIGO (2011).

\bibitem{HWS_ETM_optical}
{A. Brooks et. al.}, \enquote{{aLIGO Hartmann Sensor Optical Layouts (H1, L1,
  H2): End Test Masses},} Technical Note T1000717-v5, LIGO (2013).

\bibitem{HWS_SAD}
{A. Brooks et. al.}, \enquote{{TCS Hartmann Sensor Software Architecture},}
  Technical Note T1000155-v15, LIGO (2015).

\bibitem{RHmodel2016}
{Joshua Ramette, Marie Kasprzack, Aidan Brooks, Carl Blair, Haoyu Wang, and
  Matthew Heintze}, \enquote{{Analytical model for ring heater thermal
  compensation in the Advanced Laser Interferometer Gravitational-wave
  Observatory},} {Applied Optics} \textbf{55}, 2619--2625 (2016).

\bibitem{RH_mod_astig}
M.~Barbet and G.~Ciani, \enquote{{Status of UF Ring Heater Prototype
  development and testing},} Technical Note T1300756-v1, LIGO (2013).

\bibitem{TCS_noise_coupling}
P.~Willems, \enquote{{TCS Actuator Noise Coupling},} Technical Note T060224-v7,
  LIGO (2011).

\bibitem{CO2_mask_construction}
A.~Brooks and A.~Heptonstall, \enquote{{TCS beam shaping: optimum and
  achievable beam profiles for correcting thermo-refractive lensing (not
  thermo-elastic surface deformation)},} Technical Note T1200103-v2, LIGO
  (2012).

\bibitem{CO2_noise_coupling}
A.~Brooks, \enquote{{CO2 RIN coupling to DARM for aLIGO TCS},} Technical Note
  T1500022-v2, LIGO (2015).

\end{thebibliography}
\end{document}